\documentclass[aps,nofootinbib,superscriptaddress,floatfix]{revtex4}

\usepackage{graphicx}
\usepackage{amsmath}
\usepackage{dcolumn}
\usepackage{amsfonts}
\usepackage{amssymb}
\usepackage{dcolumn}
\usepackage{gensymb}
\usepackage{siunitx}
\usepackage{hyperref}
\begin{document}

\title{Universal spin-momentum locking of evanescent waves}

\author{Todd Van Mechelen}
\email{vanmeche@ualberta.ca}
\affiliation{Department of Electrical and Computer Engineering, University of Alberta, Edmonton T6G 2V4, Canada}

\author{Zubin Jacob}
\email{zjacob@ualberta.ca}
\affiliation{Birck Nanotechnology Center, Department of Electrical and Computer Engineering, Purdue University, West Lafayette 47906, Indiana, USA}

\date{Compiled \today}

\begin{abstract}

We show the existence of an inherent property of evanescent electromagnetic waves: spin-momentum locking, where the direction of momentum fundamentally locks the polarization of the wave.  We trace the ultimate origin of this phenomenon to complex dispersion and causality requirements on evanescent waves. We demonstrate that every case of evanescent waves in total internal reflection, surface states and optical fibers/waveguides possesses this intrinsic spin-momentum locking. We also introduce a universal right-handed triplet consisting of momentum,decay and spin for evanescent waves. We derive the Stokes parameters for evanescent waves which reveal an intriguing result - every fast decaying evanescent wave is inherently circularly polarized with its handedness tied to the direction of propagation. We also show the existence of a fundamental angle associated with total internal reflection (TIR) such that propagating waves locally inherit perfect circular polarized characteristics from the evanescent wave. This circular TIR condition occurs if and only if the ratio of permittivities of the two dielectric media exceeds the golden ratio. Our work leads to a unified understanding of this spin-momentum locking in various nanophotonic experiments and sheds light on the electromagnetic analogy with the quantum spin hall state for electrons. 

\end{abstract}

\maketitle


\section{Introduction}
\label{sec:1Introduction}


An important signature of the recently discovered quantum spin hall (QSH) state of matter is the existence of electronic surface states which are robust to disorder (non-magnetic impurities) \cite{kane2005quantum,bernevig2006quantum}. This property arises since the spin of the electron is intrinsically locked to the direction of propagation (momentum) and the  electrons cannot backscatter unless there is a spin-flip \cite{maciejko2011quantum}. Intriguingly, recent experiments have explored an analogous phenomenon in photonics showing polarization dependent directional propagation of optical modes in spontaneously emitted as well as scattered light \cite{neugebauer2014polarization,rodriguez2013near,lin2013polarization,petersen2014chiral,mitsch2014quantum,o2014spin,kapitanova2014photonic,sollner2015deterministic}. For example, experiments have shown that spontaneous emission from atomic transitions is preferentially uni-directed along a fiber depending on the magnetic quantum number of the excited state \cite{mitsch2014quantum}. On the other hand, surface plasmon polaritons excited with circularly polarized light have also demonstrated unidirectional propagation \cite{rodriguez2013near,lin2013polarization}. One common thread in these experiments is the evanescent wave which leads to a clear hint that the effect is tied to fundamental properties of decaying waves and not the details of the nanophotonic structures. A quantum field theoretic treatment has also recently shed light on the interesting spin properties of evanescent waves \cite{bliokh2014extraordinary,bliokh2012transverse}.  However, there is an urgent need for a unified theory about the inherent origin of this effect and its underlying connection to experiments. In analogy with the behavior of electrons in the quantum spin hall effect, we call this phenomenon ``spin-momentum locking".

In this paper, our central contribution is the proof that spin-momentum locking is universal behavior for electromagnetic waves which stems from the complex dispersion relation of evanescent waves and fundamental causality requirements. We introduce a universal triplet consisting of momentum, decay and spin of evanescent waves. We show that the Stokes parameters for an evanescent wave unambiguously reveals that every fast decaying evanescent wave is inherently circularly polarized irrespective of how it originates.  Furthermore, this inherent handedness (spin) is locked to the direction of propagation (momentum). This information hidden in the Stokes parameters has been overlooked till date and is in stark contrast to the existing knowledge on propagating waves.   The universality of this phenomenon is revealed by analyzing, in detail, the cases corresponding to a) total internal reflection (TIR) b) waveguides c) optical fibers d) surface electromagnetic waves. We also show the existence of a unique criterion in TIR (``golden ratio condition") at which propagating light is locally circularly polarized on total internal reflection. This effect can be used to verify our theory in near-field optical experiments. Lastly, we provide detailed insight on how spontaneous emission from a quantum emitter can couple to spin-momentum locked states in optical fibers. Our work explains various experimental observations and should open up future ways of exploiting this universal spin-momentum locking for practical applications.


\section{Evanescent Waves}
\label{sec:2ComplexkSpace}


\subsection{Complex Dispersion Relation}
\label{subsec:2.1ComplexDispersionRelation}

We first construct a general basis vector for evanescent waves independent of origin which reveals a universal electromagnetic right handed triplet consisting of momentum, decay and spin. The wavevector of an evanescent plane wave necessarily has to be complex and can be written in a general form as $\mathbf{k}=\pmb{\kappa}+i~\pmb{\eta}$. Here, $\pmb{\eta}$ is the imaginary part of $\mathbf{k}$ and is related to the decay while $\pmb{\kappa}$ is the real part related to phase propagation (momentum). Starting from the dispersion relation in free space, the square of $\mathbf{k}$ is fixed via,
\begin{equation}\label{eq:1dispersion}
\mathbf{k}^2=\mathbf{k}\cdot\mathbf{k}={k_0}^2
\end{equation}
which implies, since $k_0=\omega/c$ is purely real, that the two components of $\mathbf{k}$ must satisfy,
\begin{subequations}\label{eq:2dispersion}
	\begin{equation}
	\kappa^2-\eta^2={k_0}^2
	\end{equation}
	\begin{equation}
	\pmb{\kappa}\cdot\pmb{\eta}=0.
	\end{equation}
\end{subequations}

From Eq.~(\ref{eq:2dispersion}b), we make an important observation: the complex dispersion relation in free space necessarily requires that $\pmb{\kappa}$ and $\pmb{\eta}$ be orthogonal. This implies that the phase propagation of an evanescent wave (momentum) is perpendicular to its direction of decay. Furthermore, these orthogonal phase propagation and decay vectors always have a phase difference between them (factor of $i=\sqrt{-1}$) which are imprinted on orthogonal components of the electromagnetic field vectors through the transversality condition ($\mathbf{k}\cdot\mathbf{E}=0$). We will show now that this is the intuitive reason for the inherent handedness (spin) of the evanescent wave. 

Like propagating plane waves, evanescent waves can have two orthogonal field polarizations which we denote by $\mathbf{\hat{s}}$ and $\mathbf{\hat{p}}$ unit vectors.  $\mathbf{\hat{s}}$ is defined to have an electric field perpendicular to the plane formed by the propagation vector ($\pmb{\kappa}$) and decay vector ($\pmb{\eta}$) while the electric field vector lies in the plane for $\mathbf{\hat{p}}$. Without any loss of generality, an elegant choice of basis can be made to represent these unit vectors uniquely in terms of the evanescent wave wavevector. Our choice of basis is the triplet $\{\pmb{\kappa}$, $\pmb{\eta}$, $\pmb{\kappa}\times\pmb{\eta}\}$. We emphasize that this choice of basis alone fulfils the transversality condition imposed on electromagnetic waves in vacuum ($\mathbf{k}\cdot\mathbf{E}=0$).

By defining $\mathbf{\hat{s}}$ and $\mathbf{\hat{p}}$ as,
\begin{subequations}\label{eq:3polarization}
	\begin{equation}
	\mathbf{\hat{s}}=\frac{\pmb{\kappa}\times\pmb{\eta}}{|\pmb{\kappa}\times\pmb{\eta}|}=i~\frac{\mathbf{k}\times\mathbf{k}^*}{|\mathbf{k}\times\mathbf{k}^*|}
	\end{equation}
	\begin{equation}
	\mathbf{\hat{p}}=\frac{\mathbf{k}\times\mathbf{\hat{s}}}{|\mathbf{k}|}=i~\frac{\mathbf{k}\times(\mathbf{k}\times\mathbf{k}^*)}{|\mathbf{k}|~|\mathbf{k}\times\mathbf{k}^*|}
	\end{equation}
	\begin{equation}
	\mathbf{k}\cdot\mathbf{\hat{s}}=\mathbf{k}\cdot\mathbf{\hat{p}}=\mathbf{\hat{s}}\cdot\mathbf{\hat{p}}=0
	\end{equation}
\end{subequations}
we express the evanescent field polarization entirely in terms of its momentum ($\mathbf{k}$). This form is robust enough that it can also be generalized to lossy media when $\pmb{\kappa}$ and $\pmb{\eta}$ are non-orthogonal. We emphasize that this unique form of evanescent wave basis vectors is universal and reduces to the case of plane wave basis vectors when  $\eta\to 0$.

The above representation reveals important aspects about the intrinsic ``spin" of an evanescent wave. We define this intrinsic ``spin" to be the inherent handedness (left/right circular/elliptical polarization) of the field basis vector. We rigorously justify this in the next section but make a note that electric fields in any specific scenario can be represented using these basis vectors. Hence, properties of field basis vectors will always be manifested in the electric and magnetic fields.  

Note first that $\mathbf{\hat{s}}$ is purely real so the orthogonal  components comprising the field vector will be in phase. Thus evanescent waves with electric field vector field perpendicular to the plane formed by the decay vector and propagation vector will show no interesting polarization characteristics. However, the $\mathbf{\hat{p}}$ field basis vector is now complex. 
Using the properties from Eq.~(\ref{eq:2dispersion}) and a bit of manipulation we obtain,
\begin{equation}\label{eq:5p-polarization}
\mathbf{\hat{p}}=i\bigg[\frac{\eta}{|\mathbf{k}|}\bigg(\frac{\pmb{\kappa}}{\kappa}\bigg)+i\frac{\kappa}{|\mathbf{k}|}\bigg(\frac{\pmb{\eta}}{\eta}\bigg)\bigg]
\end{equation}
where we can clearly see that the $\mathbf{\hat{p}}$-polarization is just a linear combination of $\pmb{\kappa}$ and $\pmb{\eta}$ unit vectors with an in-built phase difference
between the orthogonal components. This immediately implies that there will be an inherent elliptical polarization imparted to the field. 

\begin{figure}
    \includegraphics{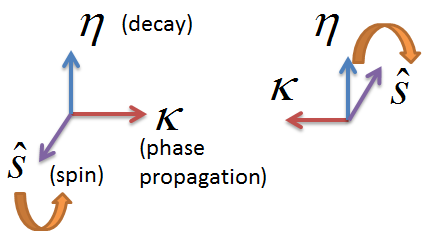}
    \caption{Our result shows a fundamental right handed triplet formed by momentum, decay and spin for evanescent waves. Note the locked triplets for waves propagating in two opposite directions. As we can see, the direction of the spin $\mathbf{\hat{s}}$ flips for the two cases. It is important to note that in general there are four degenerate solutions but two of these correspond to growing evanescent waves which are forbidden due to causality. This explains why the left handed triplet is not allowed and the phenomenon of spin-momentum locking is universal to evanescent waves.}
    \label{fig:EvanescentBasis}
\end{figure}


\subsection{Stokes Parameters}
\label{subsec:2.2StokesParameters}

We now extend the concept of Stokes parameters \cite{mcmaster1961matrix}  beyond propagating waves to fully characterize this interesting $\mathbf{\hat{p}}$-polarization state of an evanescent wave. Complex $\mathbf{\hat{p}}$ is expressed as a linear combination of two basis vectors which motivates us to consider spin-$\frac{1}{2}$ operators. The Stokes parameters of an evanescent wave can be written as the expectation values of the Pauli matrices and carries non-trivial information;
\begin{subequations}\label{eq:6stokesparamaters}
	\begin{equation}
	S_0 = \langle\mathbf{\hat{p}}| 1 |\mathbf{\hat{p}}\rangle = 1
	\end{equation}
	\begin{equation}
	S_1 = \langle \mathbf{\hat{p}}| \sigma_z |\mathbf{\hat{p}}\rangle = \frac{{k_0}^2}{|\mathbf{k}|^2}
	\end{equation}
	\begin{equation}
	S_2 = \langle \mathbf{\hat{p}}| \sigma_x |\mathbf{\hat{p}} \rangle = 0
	\end{equation}
	\begin{equation}
	{S_3}^{\pm} = \langle \mathbf{\hat{p}}| \sigma_y |\mathbf{\hat{p}} \rangle = \pm 2\frac{\kappa~\eta}{~|\mathbf{k}|^2}.
	\end{equation}
\end{subequations}

$S_1$ and $S_3$ quantify the amount of spin, i.e. the degree of linear and circular polarized character of an electromagnetic wave. Here, $\pm$ denotes the two directions of the phase propagation possible for the evanescent wave. We see that the polarization state of the field basis vector $\mathbf{\hat{p}}$ is dependent only on the complex components of the wavenumber while the actual electric and magnetic field elements are irrelevant in this instance. This means that there will be a certain degree of elliptical polarization \textit{intrinsic} to the electromagnetic field which is determined entirely by the real and imaginary components of the momentum ($\mathbf{k})$. In this sense, there will be inherent ``spin" associated with the evanescent wave since the unique basis vector $\mathbf{\hat{p}}$ itself imparts handedness to the wave.  

Note, the $\mathbf{\hat{s}}$ vector can now be interpreted as the ``spin direction" since it signifies the handedness of the electric field with $\mathbf{\hat{p}}$-polarization. This spin vector ($\mathbf{\hat{s}}$) is orthogonal to both $\pmb{\kappa}$ and $\pmb{\eta}$ which constitute the basis of $\mathbf{\hat{p}}$. Furthermore, the transformation $\pmb{\kappa}\to-\pmb{\kappa}$, for fixed decay direction ($\pmb{\eta})$, changes the handedness of $\mathbf{\hat{p}}$ (sign($S_3$)). This also flips the direction of $\mathbf{\hat{s}}$ which is consistent with an opposite direction of spin. This shows that the spin is fundamentally locked to the direction of propagation (momentum). The diagram in Fig.~(\ref{fig:EvanescentBasis}) shows the construction of a fundamentally locked triplet related to $\mathbf{\hat{p}}$-polarized evanescent waves formed by the phase propagation vector ($\pmb{\kappa}$), decay vector ($\pmb{\eta}$) and spin ($\mathbf{\hat{s}}$).

\begin{figure}
    \includegraphics{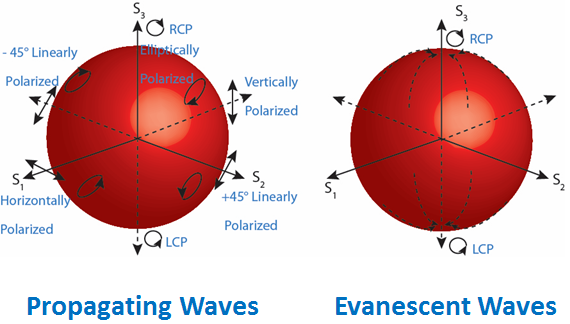}
    \caption{Poincar\'{e} spheres for propagating waves and $\mathbf{\hat{p}}$-polarized evanescent waves. Propagating waves can have any arbitrary polarization state for a given phase velocity. However, all fast decaying evanescent waves are circularly polarized and lie on the south or north pole of the Poincar\'{e} sphere ($S_3=\pm1$). Furthermore, the choice between these two points is locked to the direction of momentum ($\pm \pmb{\kappa}$).}
    \label{fig:BlochSphere}
\end{figure}


\subsection{Inherent Polarization}
\label{subsec:2.3InherentPolarizationofEvanescentWaves}

In this section, we prove that every fast decaying evanescent wave is inherently circularly polarized and its handedness is tied to the direction of phase propagation (momentum). We consider the case of an evanescent wave with very high momentum such that $\kappa \gg k_0$. The dispersion relation then implies  $\kappa \approx \eta$ and the wave decays on a length scale much smaller than the wavelength. Simplifying the expression for the  $\mathbf{\hat{p}}$-polarized basis vector,

\begin{subequations}\label{eq:10reducedp}
\begin{equation}
\mathbf{\hat{p}}\to \frac{i}{\sqrt{2}}\bigg[\bigg(\frac{\pmb{\kappa}}{\kappa}\bigg)+i\bigg(\frac{\pmb{\eta}}{\eta}\bigg)\bigg]
\end{equation}
\begin{equation}
S_1 \to 0
\end{equation}
\begin{equation}
{S_3}^{\pm} \to \pm 1
\end{equation}
\end{subequations}
which we can clearly see is a state of circular polarization.

The above result implies that every fast decaying evanescent wave lies on the north or south pole of the Poincar\'{e} sphere while propagating waves can lie anywhere on the Poincar\'{e} sphere. Furthermore, the choice of the south and north pole ($S_3=\pm1$) is dictated by the direction of the phase velocity ($\pm \pmb{\kappa}$).  Thus spin-momentum locking is a fundamental property of evanescent waves. To visually illustrate these polarization states, we compare the Poincar\'{e} spheres of propagating and $\mathbf{\hat{p}}$-polarized evanescent waves in Fig.~(\ref{fig:BlochSphere}). 



\section{Spin-momentum Locking From Causality}
\label{sec:3SpinMomentumLocking}

The ``spin-locking" characteristic of evanescent waves comes from the fact that $\pmb{\kappa}$ and $\pmb{\eta}$ are inherently orthogonal as dictated by the complex dispersion (Eq.~(\ref{eq:2dispersion})). Simultaneously, the unit field vector $\mathbf{\hat{p}}$ which is related to the wavevector possesses a $\pi/2$ phase difference between its orthogonal components. This phase is not an artifact of some particular combination of polarization vectors but is \textit{embedded into the vector field itself to guarantee that the transverse condition ($\mathbf{k}\cdot\mathbf{E}=0$) is satisfied}.

Ultimately, evanescent waves require some sort of boundary condition to exist in a region of space, which usually involves a symmetry breaking or a change in material parameters. For an arbitrary plane wave ($\propto \exp(i~\mathbf{k}\cdot\mathbf{r})=\exp(i~\pmb{\kappa}\cdot\mathbf{r})\exp(-\pmb{\eta}\cdot\mathbf{r})$), this boundary condition, in general, opens up 2 possible propagation directions $\pm \pmb{\kappa}$, and 2 decay/growth directions $\pm \pmb{\eta}$ which allows up to 4 degenerate solutions. However, we know immediately that only one of the $\pmb{\eta}$ solutions can exist because the wave must be finite in the region of space that includes infinity, i.e. it must decay away from the boundary towards infinity. Exponential growth in a passive medium is non-physical because it would require a non-causal solution to the boundary condition. 


This causality requirement leads to the fact that the handedness or ``spin" of the evanescent waves is now determined and locked to the propagation direction (the momentum). This is because while the decay vector ($\pmb{\eta}$) cannot change, the wave is free to propagate in both directions ($\pm \pmb{\kappa}$), flipping the handedness of $\mathbf{\hat{p}}$. In other words, the set of allowed evanescent waves only consists of 2 possibilities due to this condition. One with positive momentum $+\pmb{\kappa}$ and positive spin direction $+\mathbf{\hat{s}}$ and the other with negative momentum $-\pmb{\kappa}$ and negative spin direction $-\mathbf{\hat{s}}$. Hence, causality and transversality (or complex dispersion) can be considered to be the fundamental origin of the universal spin-momentum locking of evanescent waves (see Fig.~(\ref{fig:EvanescentBasis})).


\section{Universal Behavior}

In this section, we show that evanescent waves possess this spin-momentum locking in various scenarios. It becomes imperative to revisit fundamental concepts of total internal reflection and waveguide modes to prove that evanescent waves indeed possess a property which has been overlooked. To analyze these textbook phenomena, we introduce the concept of a local handedness for inhomogeneous fields. We specifically plot the spatial distribution of the Stokes parameter ($S_3$) which depends on the local electric fields and sheds light on the local handedness (polarization state) of the fields. We note that our approach is different but equivalent to the historic concept of the light beam tensor introduced by Fedorov \cite{fedorov1965covariant} and the recently developed concept of the spin density \cite{cameron2012optical,bliokh2012transverse,bliokh2013dual}.

\label{sec:4UniversalBehavior}


\subsection{Circular Total Internal Reflection (Golden Ratio Condition)}
\label{subsec:4.1CircularTIR}

\begin{figure}
    \includegraphics[width=0.75\columnwidth]{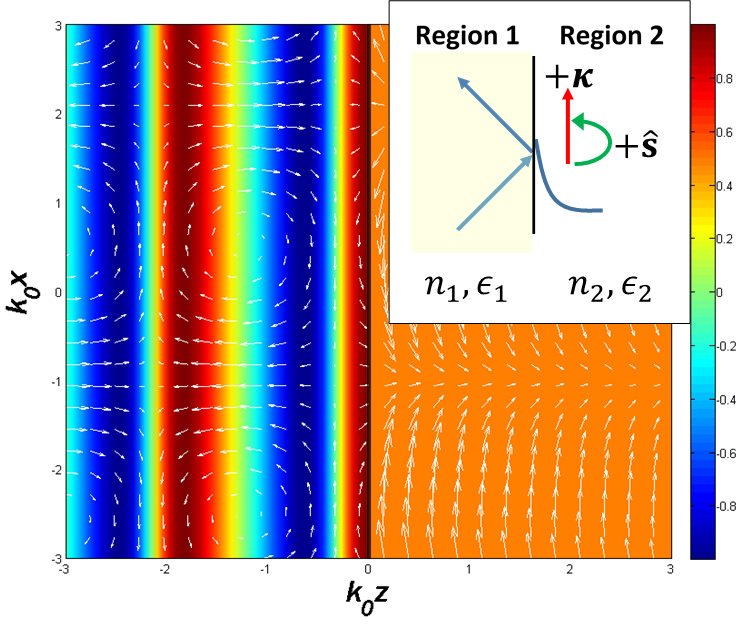}
    \caption{CTIR at interface between glass with $n_{1}=2$ and air with $n_{2}=1$ at the $\theta_{\mathrm{CTIR}}$ condition. For waves travelling in the $+x$ direction, the evanescent wave in region 2 has right handed spin-momentum locking (inset). Note the wave in medium 1 has perfect circular polarization characteristics close to the interface at this angle of incidence. The overlaid false color plot is the spatial distribution of the normalized Stokes parameter ($S_3$) which characterizes the handedness of the wave (degree of circular polarization) from $-1$ to $1$ at each point.}
    \label{fig:1TIR1}
    \includegraphics[width=0.75\columnwidth]{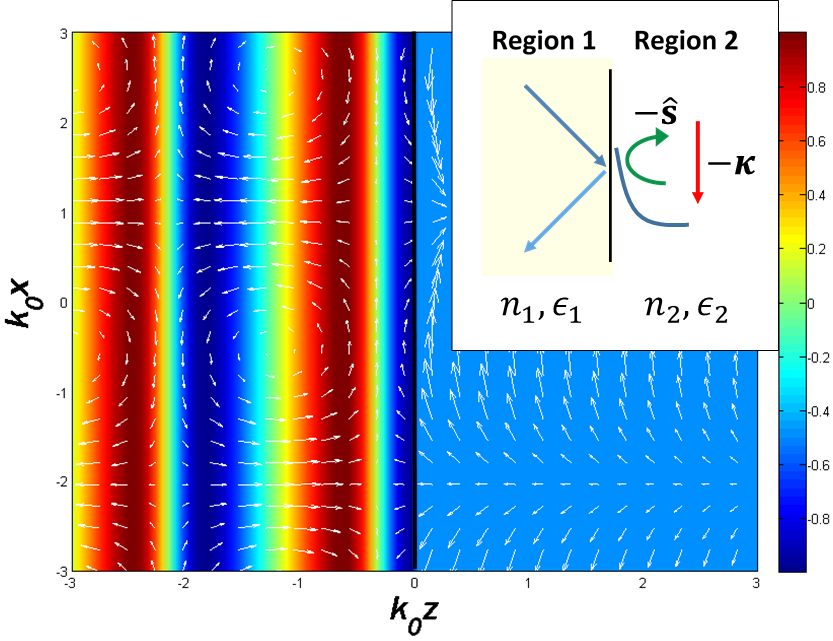}
    \caption{CTIR at interface between glass with $n_1=2$ and air with $n_2=1$ at the $\theta_{\mathrm{CTIR}}$ condition. For waves travelling in the $-x$ direction, the evanescent wave in region 2 has left handed spin-momentum locking (inset). The plot illustrates that the evanescent wave spin has the opposite sign as compared to the previous case because the momentum and spin are locked.}
    \label{fig:2TIR2}
\end{figure}

The simplest case where such a phenomenon can occur is when evanescent waves are generated at total internal reflection (TIR). We consider a wave $\mathbf{\hat{p}}$-polarized in the $\mathbf{\hat{x}}$-$\mathbf{\hat{z}}$ plane (TM) travelling from glass with index $n_{1}=\sqrt{\epsilon_1}$ into medium 2 with index $n_{2}=\sqrt{\epsilon_2}$ where we require $\epsilon_1>\epsilon_2$ for evanescent waves to be supported. The electric fields generated during TIR are well known and are depicted by white arrows in Fig.~(\ref{fig:1TIR1}) and (\ref{fig:2TIR2}). However, when overlaid against the local handedness of the field  an intriguing phenomenon comes to light - the direction of propagation of the wave alters the relative handedness of the evanescent field. The false colors in the same figures depict the spatial distribution of the normalized Stokes parameter ($S_3$) and quantifies the polarization state of the field at each point. In region 2, it is evident that the evanescent wave possesses similar handedness at every point (orange region). Furthermore, comparing the counter-propagating cases between Fig.~(\ref{fig:1TIR1}) and (\ref{fig:2TIR2}) we clearly see that the polarization state of the evanescent wave is reversed and the Stokes parameter changes sign. The insets of Fig.~(\ref{fig:1TIR1}) and (\ref{fig:2TIR2}) elucidate this spin-momentum locking phenomenon for TIR.

We now show that the propagating waves inherit handedness from the evanescent waves due to boundary conditions at the interface.
The phase between the perpendicular and parallel components of an arbitrary ($\mathbf{\hat{p}}$-polarized) electric field in region 1, interfaced with an evanescent wave in region 2 must satisfy
\begin{equation}\label{eq:13electricfieldratio}
\bigg[\frac{E_{\perp}}{E_{\parallel}}\bigg]_{1}=\pm i~\frac{\epsilon_{2}}{\epsilon_{1}}~\bigg[\frac{\kappa}{\eta}\bigg]_{2} \qquad @~\textrm{interface}
\end{equation}
where the $\pm$ indicates oppositely travelling evanescent waves and the subscripts designate the field components in their respective material regions. It should be stressed that this only applies \textit{locally} at the interface. However, this could have interesting consequences for near-field optics since it implies that there is a preferential handedness depending on the direction of propagation when we couple to evanescent waves. 
We make the important observation that perfect circular polarization is enforced (locally) in region 1 when
\begin{equation}\label{eq:14perfectcp}
\frac{\epsilon_{1}}{\epsilon_{2}}=\bigg[\frac{\kappa}{\eta}\bigg]_{2}.
\end{equation}
We can now solve for the momentum and decay of the evanescent wave which achieves this circular total internal reflection. They are
\begin{equation}
\kappa_2=\epsilon_1\sqrt{\frac{\epsilon_2}{{\epsilon_1}^{2}-{\epsilon_{2}}^2}}k_0
\end{equation}
and
\begin{equation}
\eta_2=\epsilon_2\sqrt{\frac{\epsilon_2}{{\epsilon_1}^{2}-{\epsilon_{2}}^2}}k_0.
\end{equation}
In the case of TIR, this local circular polarization is generated in region 1 because there is a phase shift imparted to the reflected wave and the interference with the incident wave causes the combined field to be locally handed. Lastly, we need to determine the angle of incidence of the propagating wave that is required to accomplish this circular TIR condition. Using Snell's law, it can be shown that the CTIR angle of incidence $\theta_1=\theta_{\mathrm{CTIR}}$ is,
\begin{equation}\label{ctirangle}
\sin(\theta_{\mathrm{CTIR}})=1/\sqrt{\epsilon_{1}/\epsilon_{2}-\epsilon_{2}/\epsilon_{1}}.
\end{equation}
We note that in this instance, $\theta_{\mathrm{CTIR}}$ must necessarily be real which requires that $\sqrt{\epsilon_{1}/\epsilon_{2}-\epsilon_{2}/\epsilon_{1}}>1$. Therefore, there is an interesting limiting condition for local CTIR to exist which is when $\theta_{\mathrm{CTIR}}\to\pi/2$ (i.e. when the propagating wave in region 1 is parallel to the interface). This is equivalent to the limit when
\begin{equation}
\bigg[\frac{\epsilon_{1}}{\epsilon_{2}}\bigg]_{\mathrm{GR}}=\frac{1}{2}(1+\sqrt{5})\approx 1.618
\end{equation}
which is the \textit{minimum} allowable ratio of the permittivities for CTIR to occur, and curiously, it can also be identified as the golden ratio \cite{livio2008golden}. We term this the ``golden ratio condition" for local circularly polarized total internal reflection. 

This induced CTIR in region 1 is visible clearly in Fig.~(\ref{fig:1TIR1}) and (\ref{fig:2TIR2}). Note our choice of refractive indices satisfies $\epsilon_1/\epsilon_2=4 > [\epsilon_1/\epsilon_2]_{\mathrm{GR}}$. The angle given by our analytical expression in Eq.~(\ref{ctirangle}) is $\theta_{\mathrm{CTIR}}=31.09^{\mathrm{o}}$ and we have plotted the fields for this incident angle. Close to the interface in region 1, the Stokes parameter takes the maximal values of $S_3=\pm 1$ (red and blue regions). Thus the fields are perfectly circular polarized close to the interface specifically for this angle of incidence. Although phase propagation normal to the interface ($\hat{\mathbf{z}}$) will alter the degree of this polarization, the \textit{relative} handedness between forward and backward propagating waves is maintained. This can be seen from the blue and red contours in region 1, where rotation of the electric field vectors is reversed at every point in space - which is in agreement of differing signs of $S_3$.


\subsection{Waveguides}
\label{subsec:4.3Waveguides                                                                                                                                                                                                                                                                                                                                                                                                                                                                                                                                                                                                                                                                                                                                                                                                                                                                                                                                                                                                                                                                                                                                                                                                                                                                                                                                                                                                                                                                                                                                                                                                                                                                                                                                                                                                                                                                                                                                                                                                                                                                                                                                                                                                                                                                                                                                                                                                                                                                                                                                                                                                                                                                                                                                                                                                                                                                                                                                                                                                                                                                                                                                                                                                                                                                                                                                                                                                                                                                                                                                                                                                                                                                                                                                                                                                                                                                                                                                                                                                                                                                                                                                                                                                         }                                                                                                                                                                                                                                                                                                                                                                                                                                                                                                                                                                                                                                                                                                                                                                                                                                                                                                                                                                                                                                                                                                                                                                                                                                                                                                                                                                                                                                                                                                                                                                                                                                                                                                                                                                                                                                                                

\begin{figure}
    \includegraphics[width=0.75\columnwidth]{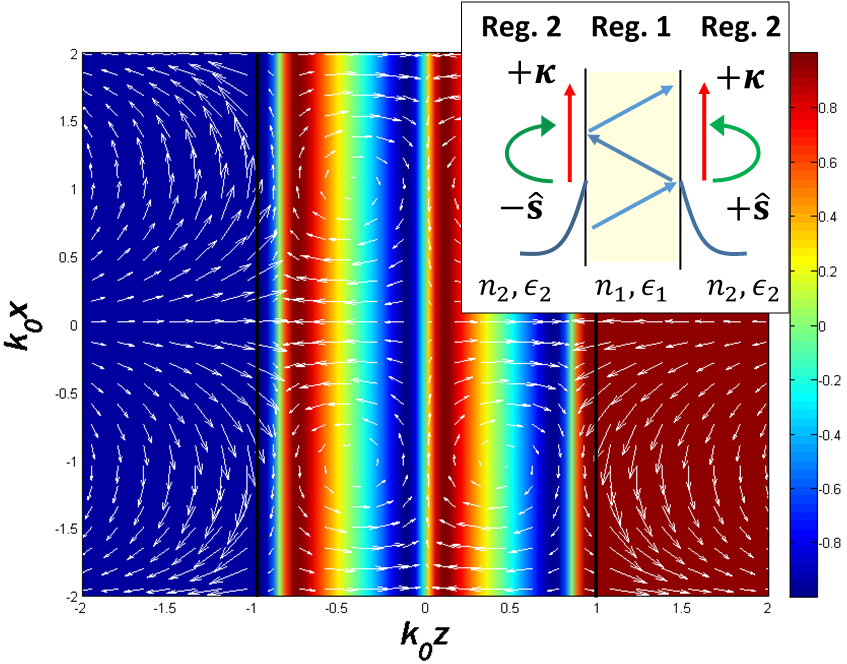}
    \caption{Waveguide mode at interface between glass with $n_1=4$ and air with $n_2=1$. The width of the waveguide is $2k_{0}d=2$. For waveguide modes travelling in the $+x$ direction, the evanescent waves in region 2 lock the handedness (locally) to $+\mathbf{\hat{s}}$ at $k_{0}z=1$ and $-\mathbf{\hat{s}}$ at $k_{0}z=-1$. The false color plot shows the spatial distribution of the normalized Stokes parameter ($S_3$) from $-1$ to $1$ for the waveguide and illustrates the intrinsic handedness of the evanescent waves. Furthermore, on comparison with the counter-propagating waveguide mode, we see that the handedness is reversed.}
    \label{fig:3WG1}
    \includegraphics[width=0.75\columnwidth]{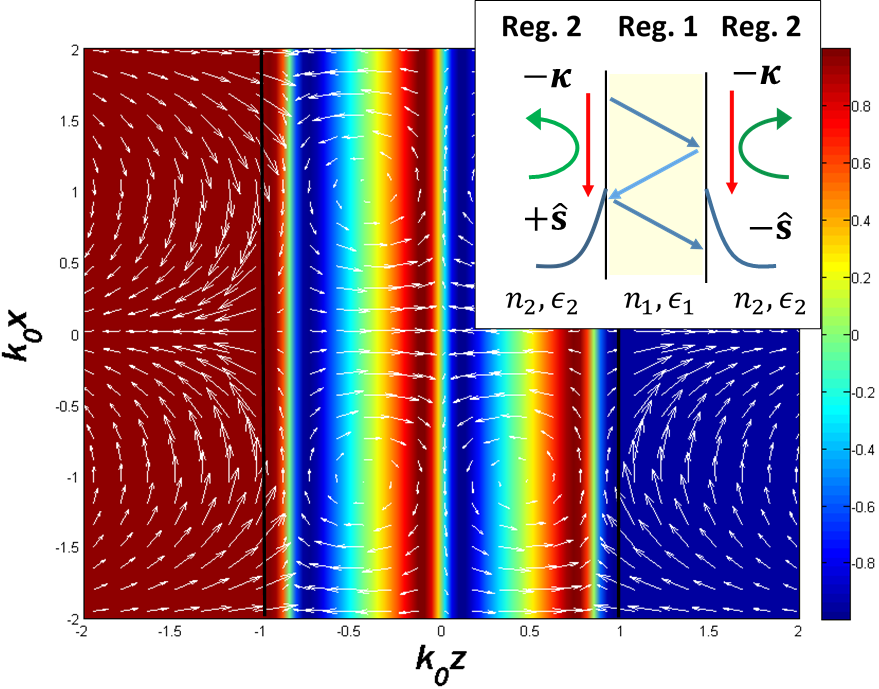}
    \caption{Waveguide mode at interface between glass with $n_1=4$ and air with $n_2=1$. The width of the waveguide is $2k_{0}d=2$. For waveguide modes travelling in the $-x$ direction, the evanescent waves in region 2 lock the handedness (locally) to $-\mathbf{\hat{s}}$ at $k_{0}z=1$ and $+\mathbf{\hat{s}}$ at $k_{0}z=-1$.}
    \label{fig:4WG2}
\end{figure}

Interesting spin-locking phenomena also occur when we consider confined light in waveguides and optical fibres. The confinement of light necessarily requires evanescent waves to be present which implies that there will be handedness imparted on the waveguide and fibre modes through the boundary conditions. For planar waveguides, there are even and odd solutions and the $\mathbf{\hat{p}}$-polarized electric field components (TM modes) inside the waveguide are proportional to
\begin{equation}\label{eq:11planarwaveguides}
\mathbf{E}\propto \bigg[k_z\bigg\{\begin{array}{c}
\sin(k_z z)\\
-\cos(k_z z)\end{array}\bigg\}\mathbf{\hat{x}}+ik_x\bigg\{\begin{array}{c}
\cos(k_z z)\\
\sin(k_z z)\end{array}\bigg\}\mathbf{\hat{z}}\bigg]e^{ik_x x}
\end{equation}
where the array inside the braces indicates the two separate solutions. Note that the electric field components along the x- and z-axis have a phase difference between them dictated solely by the boundary conditions.  If we consider a wave propagating in the opposite direction, i.e. change $k_x \to -k_x$ the wave changes handedness. We see that there is spin-momentum locking in waveguides  since $k_x$ now constitutes the momentum and also controls the relative phase between the orthogonal field components. The electric field vector plots in Fig.~(\ref{fig:3WG1}) and (\ref{fig:4WG2}) are overlaid on the spatial distribution of the $S_3$ Stokes parameters (false color plot) to illustrate the different spin (handedness) between two oppositely propagating waveguide modes. We note that a similar explanation can be extended to the case of  metamaterials \cite{kapitanova2014photonic}. This is discussed briefly in the supplementary information and a detailed derivation will be presented elsewhere. 

\subsection{Optical Fibres}
\label{subsec:4.4OpticalFibres}

We now show that spin-momentum locking in optical fibres is the fundamental origin of recent experimental observations where scattered light and spontaneous emission was directed preferentially along the fiber \cite{mitsch2014quantum,petersen2014chiral}. The $\mathrm{HE_{11}}$ fundamental mode operation is the most important  case  so we  quantify its degree of polarization. To characterize our fibre mode we consider weakly guided waves, $\Delta=({n_1}^2-{n_2}^2)/(2{n_1}^2)\approx (n_1-n_2)/n_1 \ll 1$ with a numerical aperture, $\mathrm{NA}=\sqrt{{n_1}^2-{n_2}^2}\approx n_1 \sqrt{2\Delta}$~. For single mode $\mathrm{HE_{11}}$ operation, we require that $\mathrm{V}=2\pi(a/\lambda_0)\mathrm{NA}=\sigma_1 \sqrt{2\Delta}<2.405$, where $a$ is the radius of the fibre and $\sigma_1=k_1 a=2 n_1 \pi(a/\lambda_0)$ is the scaling parameter inside the core.


The $\mathrm{HE_{11}}$ is doubly degenerate in that we have two counter-rotating angular momentum modes in the plane perpendicular to the fiber-optic axis. We denote the electric and magnetic fields as $\mathbf{E}_{m}$ and $\mathbf{H}_{m}$ respectively where the subscripts denote $m=+1$ or $m=-1$. In the circular basis we define unit vectors $\mathbf{\hat{e}}_{m}=(\mathbf{\hat{r}}+im~\pmb{\hat{\phi}})/\sqrt{2}$ and clearly $\mathbf{\hat{e}_{-}}=\mathbf{\hat{e}_{+}}^*$. With a propagation factor of $\exp[i(\beta~z/a-\omega t)]$ omitted, the electric and magnetic fields can then be written as,
\begin{subequations}\label{eq:HE11Inside}
\begin{equation}
\mathbf{E}_{m}=E_0[\sqrt{2}\beta J_{0}(\mathrm{X}~r/a)\mathbf{\hat{e}}_{m}+i~\mathrm{X}J_{1}(\mathrm{X}~r/a)\mathbf{\hat{z}}]e^{im\phi}
\end{equation}
\begin{equation}
\mathbf{H}_{m}=-imH_0[\sqrt{2}(\sigma_{1})^2 J_{0}(\mathrm{X}~r/a)\mathbf{\hat{e}}_{m}+i~\beta\mathrm{X}J_{1}(\mathrm{X}~r/a)\mathbf{\hat{z}}]e^{im\phi}
\end{equation}
\end{subequations}
for fields inside the fibre when $r<a$, where $H_0=E_0/(\omega\mu_{0}a)$ and
\begin{subequations}\label{eq:HE11Outside}
\begin{equation}
\mathbf{E}_{m}=\mathcal{N}E_0[\sqrt{2}\beta K_{0}(\mathrm{Y}~r/a)\mathbf{\hat{e}}_{m}+i~\mathrm{Y}K_{1}(\mathrm{Y}~r/a)\mathbf{\hat{z}}]e^{im\phi}
\end{equation}
\begin{equation}
\mathbf{H}_{m}=-im\mathcal{N}H_0[\sqrt{2}(\sigma_{2})^2 K_{0}(\mathrm{Y}~r/a)\mathbf{\hat{e}}_{m}+i~\beta\mathrm{Y}K_{1}(\mathrm{Y}~r/a)\mathbf{\hat{z}}]e^{im\phi}
\end{equation}
\end{subequations}
outside the fibre when $r>a$ and $\mathcal{N}=(\mathrm{X}/\mathrm{Y})J_{1}(\mathrm{X})/K_{1}(\mathrm{Y})$. $J_{n}$ and $K_{n}$ are the Bessel and Modified Bessel functions of order $n$ respectively. The normalized propagation constants are defined as, $|\beta|=\sqrt{(\sigma_1)^2-\mathrm{X}^2}=\sqrt{(\sigma_2)^2+\mathrm{Y}^2}$ and $\mathrm{V}^2=\mathrm{X}^2+\mathrm{Y}^2$. The components of the $\mathbf{E}_{m}$ and $\mathbf{H}_{m}$ fields have identical forms (up to a proportionality constant) so we concentrate on the electric type.

The above equations are commonplace in textbooks on fiber optics. However, the differentiation between the angular momentum and spin components of the $\mathrm{HE_{11}}$ mode has not been done before. This can be done unambiguously by extending our concept of local handedness of a wave to three dimensions. We consider the Stokes parameter ($S_3$) which characterizes circular polarization. However, for the optical fiber, it has to be evaluated for three dimensional fields by considering pairs of orthogonal directions. This leads to Stokes parameters $S_3^z$ and $S_3^{\phi}$ which can be interpreted as local circular polarization of the field with handedness along the $\mathbf{\hat{z}}$ direction or $\pmb{\hat{\phi}}$ direction. We concentrate on the field components inside the core when $r<a$, but similar expressions hold for $r>a$ where the Bessel functions are substituted with the Modified Bessel functions.

For the two $m=\pm1$ angular momentum modes, the $S_3^z$ Stokes parameter evaluated with electric field components orthogonal to the propagation $\mathbf{\hat{z}}$ direction is
\begin{equation}\label{eq:CircularIntensity}
(I_{AM})_{m}=m~2|E_0|^2\beta^2 {J_0}^2(\mathrm{X}~r/a)
\end{equation}
which we denote as the angular momentum intensity. The handedness of this angular momentum is either positive or negative for the $m=\pm 1$ modes. This is valid even if we change the sign of the propagation constant, i.e. if the $\mathrm{HE_{11}}$ mode moves along $-\mathbf{\hat{z}}$. Thus, both forward and backward propagating waves can have either positive or negative angular momentum as is expected.

However, a fundamental and intriguing asymmetry is noticed for the $S_3^{\phi}$ Stokes parameter evaluated using electric field components orthogonal to $\pmb{\hat{\phi}}$. It is given by the expression
\begin{equation}
(I_S)_{m}=\mathrm{sign}(\beta)~2|E_0|^2|\beta|\mathrm{X}J_{0}(\mathrm{X}~r/a)J_{1}(\mathrm{X}~r/a)
\end{equation}
which we denote as the spin polarization intensity. The direction of this ``spin" is in the unique $\pmb{\hat{\phi}}$ direction and is seen to be independent of the sign of the angular momentum. Furthermore, it is also locked to the momentum $\beta$ since $\mathrm{sign}(\beta)=\pm 1$ leading to fundamentally different behavior of forward and backward propagating $\mathrm{HE_{11}}$ modes along the fiber. For forward momentum $\mathrm{sign}(\beta)=+1$ we have $+\pmb{\hat{\phi}}$ transverse spin and for $\mathrm{sign}(\beta)=-1$ we have $-\pmb{\hat{\phi}}$ regardless of which angular momentum mode we are considering. Therefore, instead of four degenerate solutions, only two are allowed.

We emphasize once again that the spin-momentum locking arises from the fact that growing solutions for evanescent waves outside the optical fiber are discarded. These growing solutions have the opposite spin direction for a given propagation direction. (Sec.~\ref{sec:3SpinMomentumLocking}). This shows we have spin-momentum locking even in standard optical fibres which is directly linked to the evanescent fields necessary for confinement. Strictly speaking, we enforced spin-momentum locking from the outset by only permitting $K_n$ type Modified Bessel functions and discarding the $I_n$ type - since they exponentially grow as $r$ increases. This causality requirement with regards to fiber modes is the precise reason that we have handedness imparted to the optical fiber. 

\begin{figure}
    \includegraphics[width=0.75\columnwidth]{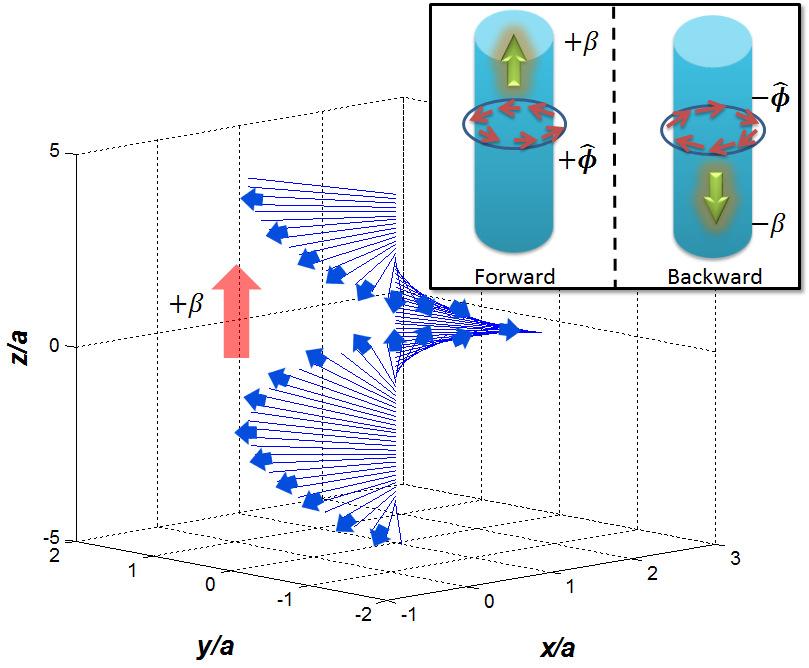}
    \caption{The evolution of the polarization vector as it propagates in an optical fibre with $\mathrm{V}=1.5$ and $\Delta=0.1$. We display the electric field at a single point at $r=a$ in the $m=+1$ $\mathrm{HE_{11}}$ mode to demonstrate the transverse spin near the core-cladding region. As we can see, the electric field rotates in the z-plane as well as in the x-y plane, hence there is a spin component directed around $\pmb{\hat{\phi}}$ (inset). Out of four possible degenerate solutions, only two are allowed because of the decaying condition on evanescent waves outside the core. Consequently, the $\mathrm{HE_{11}}$ mode of the optical fiber has spin-momentum locking.}
    \label{fig:HE11Transverse}
\end{figure}

\begin{figure}
    \includegraphics{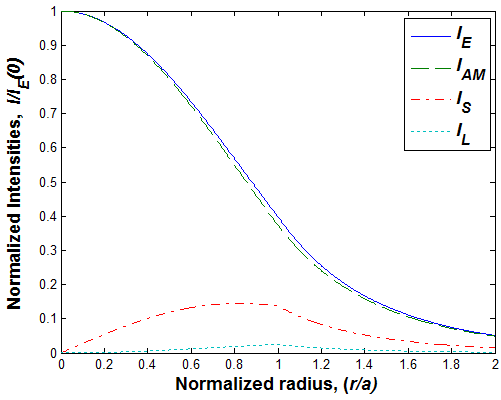}
    \caption{Normalized $\mathrm{HE_{11}}$ polarization intensities ($I/I_E(0)$) for an optical fibre of $\mathrm{V}=1.5$ and $\Delta=0.1$. We see that the majority of field is concentrated in the $I_{AM}$ angular momentum component but there is a significant component of  spin intensity ($I_S$) in the $\pmb{\hat{\phi}}$ direction near the core-cladding interface at $r=a$.}
    \label{fig:HE11Polarizations}
\end{figure}

The total electric field intensity is a sum of linear, angular momentum and spin intensities which arises from the properties of Stokes parameters (${S_0}^2 = {S_1}^2 + {S_2}^2 + {S_3}^2$). We thus have 
${I_E}^2={I_{AM}}^2+{I_S}^2+{I_L}^2$ where $I_E=|\mathbf{E}|^2=2|E_0|^2\beta^2{J_{0}}^2+|E_0|^2\mathrm{X}^2{J_{1}}^2$ is the total intensity of the electric field. Here, the linear polarization intensity is defined as $I_L=|E_0|^2\mathrm{X}^2{J_{1}}^2$, arising due to the electric field component in the $\mathbf{\hat{z}}$ direction. We can now analyze the fractional field intensity residing in the angular momentum, spin or linear polarization components.  The normalized polarization intensities for a weakly-guiding optical fibre are shown in the plot of Fig.~(\ref{fig:HE11Polarizations}). We also include a field vector plot in Fig.~(\ref{fig:HE11Transverse}) to help visualize the transverse spin component in the $\mathrm{HE_{11}}$ mode. 


\subsection{Directional Quantum Emitter Coupling}
\label{subsec:4.5quantum-emitter}

All that being said, this intriguing symmetry breaking could be exploited for applications in the field of quantum photonics. One recent experiment has utilized cold atoms near optical fibers to demonstrate directional waveguiding  of spin-polarized spontaneous emission\cite{mitsch2014quantum}. We show how this phenomenon is related to spin-momentum locking of the $\mathrm{HE_{11}}$ mode. Note, our results can be expanded to an isotropic scatterer with circularly polarized incident light or a chiral scatterer  with linearly polarized incident light.


Let us consider a left and right handed circularly polarized source that has both electric and magnetic moments. Following the semiclassical theory of spontaneous emission \cite{ford1984electromagnetic,klimov2012engineering,lee2012role}, we approximate this chiral source to be
\begin{equation}\label{eq:CurrentDensities}
\left [ \begin{array}{c}
\textbf{J}_\textsc{E}(\textbf{r}) \\
\textbf{J}_\textsc{H}(\textbf{r}) \end{array} \right]_{\pm} =
-i~\omega~\delta^{3}(\textbf{r}-\textbf{r}_0)\left [ \begin{array}{c}
\mathbf{p} \\
\mathbf{m} \end{array} \right]\\ 
=-i~\omega~\delta^{3}(\textbf{r}-\textbf{r}_0)\left [ \begin{array}{c}
p_0 \\
-i~m_0 \end{array} \right]\mathbf{\hat{e}_{\pm}}~e^{\pm i \phi}
\end{equation}
where the $\pm$ indicates left or right handed circular polarization in the cylindrical coordinate basis of the optical fiber. The coupling strength (energy of interaction) into one of the $\mathrm{HE_{11}}$ modes is then proportional to $A_{m} \propto  i\omega[\mathbf{p}^*\cdot\mathbf{E}_{m}(\mathbf{r}_0)+\mathbf{m}^*\cdot\mathbf{H}_{m}(\mathbf{r}_0)$]. Plugging in for $\mathbf{r}_0=\mathbf{0}$ it can be shown that the magnitude of the coupling strength ($|A_{m}|^2$) for each $m=\pm 1$ mode is equal to
\begin{equation}\label{eq:ChiralCoupling}
|A_{m}|^2 = C_1\bigg|\mathrm{sign}(\beta)|\beta|\omega p_0 + m\frac{(\sigma_1)^2 m_0}{\mu_0 a}\bigg|^2
\end{equation}
where $C_1$ is some positive proportionality constant.  The angular momentum quantum number $m=\pm 1$ should not be confused with the magnitude of the magnetic dipole $|\mathbf{m}|=m_0$.  Also note, that the time averaged power along the fiber axis for each mode is proportional to $P_m\propto |A_{m}|^2$. 

We notice the striking fact that this coupling factor of the chiral emitter into the $\mathrm{HE_{11}}$ mode is direction dependent. The $+$ polarization chiral emitter couples only into the $m=+1$ mode and emits most strongly in the forward propagating $\mathrm{sign}(\beta)=+1$ direction while being weaker for backward propagation $\mathrm{sign}(\beta)=-1$. Conversely, the $-$ polarization chiral emitter couples only into the $m=-1$ mode and emits more strongly in the  $\mathrm{sign}(\beta)=-1$ direction rather than $\mathrm{sign}(\beta)=+1$. This means we can control the directional propagation of waves \textit{and} the specific angular momentum mode ($m=\pm 1$) we couple into by choosing either left or right handed chiral emitters. This effect is maximal when the electric and magnetic dipole moments are tuned to have $|\beta|\omega p_0 = \frac{(\sigma_{1})^2}{\mu_{0}a}m_0$. For weakly guided waves, $|\beta|\approx\sigma_{1}$, and it can be shown that maximal coupling will occur when $m_0 \approx Z_1 p_0$ where $Z_1=Z_0/n_1=\sqrt{\mu_0/\epsilon_1\epsilon_0}$ is the wave impedance inside the fibre. 

\begin{figure}
    \includegraphics[width=0.75\columnwidth]{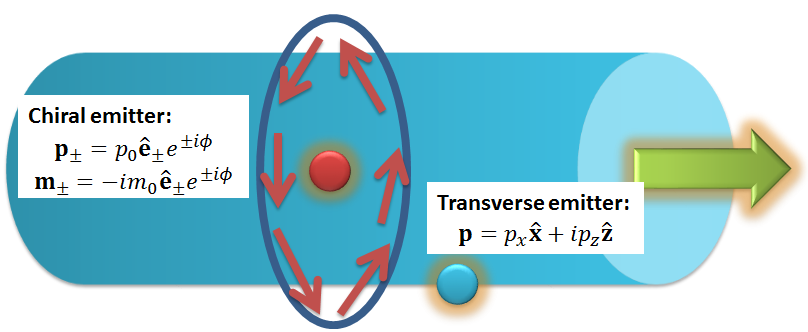}
    \caption{Chiral emitter placed at $\mathbf{r}_0=0$ and transverse emitter placed at $\mathbf{r}_0=a\mathbf{\hat{x}}$ inside the optical fibre. The intrinsic chirality of the $\mathrm{HE_{11}}$ mode opens possibilities for spin-controlled quantum photonics. We emphasize that this intrinsic chirality is universal and arises from the evanescent waves outside the core.}
    \label{fig:QE}
\end{figure}

We now propose an approach to couple strictly to the transverse spin components of the electric field with a transversely polarized electric source. This can have the advantage of not requiring magnetic dipoles or chirality. We achieve this by tuning the phase difference between two orthogonally oriented point dipole emitters $\mathbf{p}=p_x\mathbf{\hat{x}}+i~p_z\mathbf{\hat{z}}$. This emitter is placed at the location $\mathbf{r}_0=a\mathbf{\hat{x}}$ where the spin intensity is maximum (see Subsec.~\ref{subsec:4.4OpticalFibres}). The transverse spin is unchanged between angular momentum modes so they will both contribute to the propagation of the wave. The transverse coupling strength for both $m=\pm 1$ is equal to
\begin{equation}\label{eq:TransverseCoupling}
|A_m|^2 = C_2~\omega^2\bigg|\mathrm{sign}(\beta)|\beta| J_{0}(\mathrm{X})p_x+\mathrm{X}J_{1}(\mathrm{X})p_z\bigg|^2
\end{equation}
where $C_2$ is another positive proportionality constant and we see again that there is dominance of the wave to be in the $\mathrm{sign}(\beta)=+1$ direction compared to the $\mathrm{sign}(\beta)=-1$. The asymmetry in coupling between the two directions is maximal when the dipole strengths are adjusted to have $|\beta| J_{0}(\mathrm{X})p_x=\mathrm{X}J_{1}(\mathrm{X})p_z$. We illustrate these two unique quantum emitter couplings in Fig.~(\ref{fig:QE}) and their orientation in the optical fiber.

\subsection{Surface States}
\label{subsec:4.6SurfaceStates}

The last example is that of surface electromagnetic waves such as Zenneck waves \cite{jeon2006thz}, Dyakonov waves \cite{d1988new} and surface plasmon-polaritons (SPPs) which exist at the interface of two materials. The necessarily evanescent nature of the electromagnetic field will introduce very clear spin-momentum locking in all these waves. We emphasize that such polarization dependent transport has been observed for the particular case of surface plasmon polaritons \cite{rodriguez2013near,lin2013polarization,o2014spin,bliokh2012transverse} but the universality and fundamental origin of the phenomenon has never been pointed out.  

\begin{figure}
    \includegraphics[width=0.75\columnwidth]{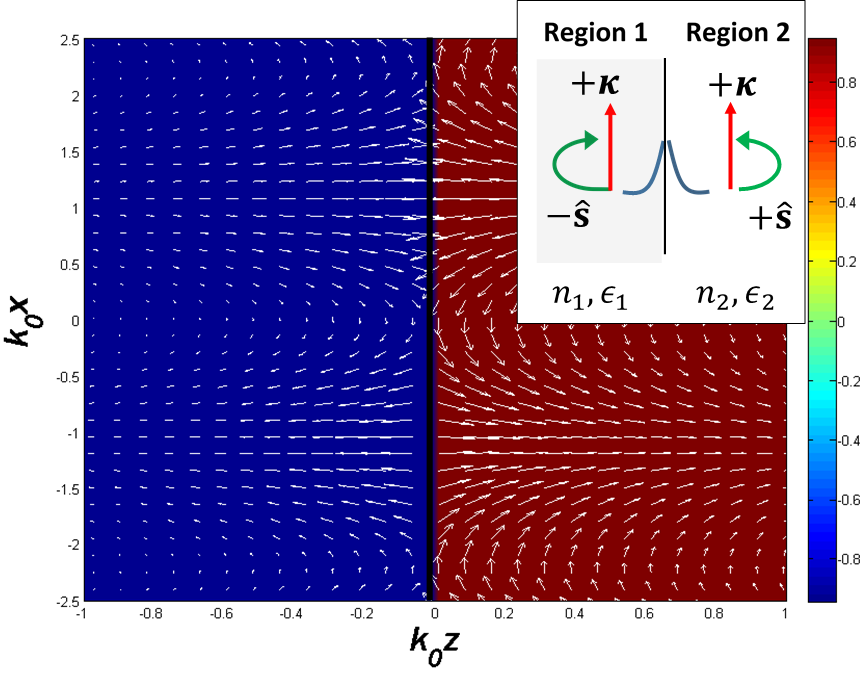}
    \caption{All electromagnetic surface waves will show spin-momentum locking. We depict here an SPP excitation between metal with $\epsilon_1=-2$ and air with $\epsilon_2=1$ propagating in the $+x$ direction. The vector plot overlaid on the spatial distribution of the Stokes parameter ($S_3$) illustrates the inherent handedness of the two evanescent waves and how they couple with counter rotating spins.}
    \label{fig:6SPP}
    \includegraphics[width=0.75\columnwidth]{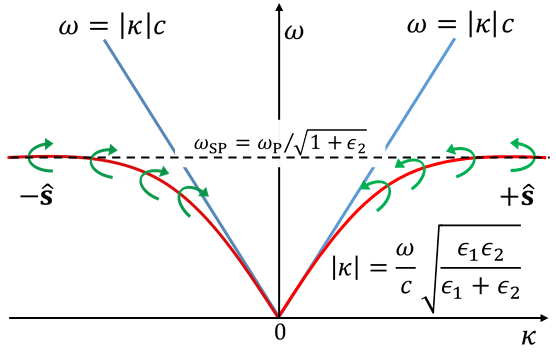}
    \caption{SPP dispersion relation that also includes the handedness of the evanescent spin (in the dielectric region). As the momentum $\kappa$ increases, the SPP spin approaches perfect circular polarization (SPP resonance).}
    \label{fig:7SPPDisp}
\end{figure}

Note that surface waves are evanescent in both regions (see Fig.~(\ref{fig:6SPP})) and hence will have \textit{global} spin-locking where the handedness of the wave will be invariant in each of the half-spaces. We explain this by taking the example of surface plasmon polaritons which exist at the interface of a metal and dielectric. Region 1 ($-z$) is metallic having a relative permittivity $\epsilon_1<0$ and the dielectric in region 2 ($+z$) has a relative permittivity $\epsilon_2>1$. This results in the familiar dispersion relation $\kappa=k_{0}\sqrt{\epsilon_{1}\epsilon_{2}/(\epsilon_{1}+\epsilon_{2})}$.
We can now fully quantify the evanescent spin in terms of the permittivities. Utilizing the expression for the circular Stokes parameters ($S_{3}$) derived in Eq.~(\ref{eq:6stokesparamaters}) this leads to
\begin{equation}
-(S_{3})_{1}=(S_{3})_{2}=2\frac{\sqrt{|\epsilon_{1}|~\epsilon_{2}}}{|\epsilon_{1}|+\epsilon_{2}}
\end{equation}
where $(S_{3})_{1}$ and $(S_{3})_{2}$ are the $\mathbf{\hat{p}}$-polarization Stokes parameters in region 1 and 2 respectively and we are assuming the permittivities are purely real in this instance. As we can see, as $|\epsilon_{1}|\to\epsilon_{2}$, the momentum $\kappa\to\infty$ and the spin approaches perfect circular polarization $-(S_{3})_{1}=(S_{3})_{2}\to1$, as expected. Also to reiterate, the spin-momentum locking of evanescent waves means these spins are flipped when the wave is propagating in the opposite direction. To help visualize these phenomena, the electric field vector plot for an SPP is displayed in Fig.~(\ref{fig:6SPP}) along with the ``full" SPP dispersion relation in Fig.~(\ref{fig:7SPPDisp}) that includes the handedness of the spin (in the dielectric region).   Our approach provides an intuitive explanation of this phenomenon observed in recent experiments where chiral emitters or near-field interference from electric and magnetic dipoles lead to unidirectional SPP propagation \cite{rodriguez2013near,lin2013polarization,o2014spin,lee2012role}.

\section{Conclusion}
\label{sec:5Conclusion}

In conclusion, we have shown that evanescent waves possess  inherent local handedness (spin) which is tied to their phase velocity (momentum). We have proven this spin-momentum locking is universal behavior since it arises due to causality and the complex dispersion relation of evanescent waves. It is interesting to note that recent work on topological photonics \cite{lu2014topological,khanikaev2013photonic,rechtsman2013photonic,gao2015topological} has shed light on the existence of surface states immune to disorder and our work will surely lead to a better understanding of those surface states as well. The QSH surface state has electrons with spins locked to their direction of propagation but only occurs on the surface (interface) of materials with spin-orbit coupling (eg: HgTe quantum wells). The electromagnetic surface state curiously always possesses this property irrespective of the nature of the material. This warrants a deeper investigation and simultaneously opens up possibilities for practical applications.

\section*{Funding Information}

We acknowledge funding from Helmholtz Alberta Initiative and National Science and Engineering Research Council of Canada.




\bibliography{References_spin_momentum_locking}


\end{document}